\pdfoutput=1
\documentclass[11pt]{article}
\usepackage[pdftex]{graphicx,color} 
\usepackage{jheppub}
\usepackage{amsmath}
\usepackage{amssymb}
\usepackage{multirow}
\usepackage{mathtools}
\usepackage{comment}
\usepackage{dsdshorthand}
\usepackage[utf8]{inputenc}
\usepackage{subcaption}
\DeclareMathOperator{\csch}{csch}
\DeclareMathOperator{\sech}{sech}
\newcommand{\bea}{\begin{equation}\begin{aligned}}
\newcommand{\eea}[1]{\label{#1}\end{aligned}\end{equation}}
\newcommand{\boa}{\begin{align}}
\newcommand{\eoa}{\end{align}}
\newcommand{\beq}{\begin{equation}}
\newcommand{\eeq}{\end{equation}}

\def\D{\Delta}
\def\o{\omega}
\def\ka{\mathcal{\kappa}}

\def\f{\phi}

\def\e{\eta}

\def \O{\mathcal{O}}
\def\X{\mathbf{X}}

\def\k{\mathbf{k}}
\def \p {\frac{\pi}{2}}

\newcommand{\nn}{\nonumber\\}

\title{Bulk reconstruction and Bogoliubov transformations in AdS$_2$}
\author[a]{Parijat Dey}
\author[b]{and Nirmalya Kajuri}
\affiliation[a]{Department of Physics and Astronomy,
	Uppsala University,\\
	Box 516,
	SE-751 20 Uppsala,
	Sweden}
\affiliation[b]{Department of Physical Sciences, Indian Association of Cultivation of Science,\\ Jadavpur, Kolkata 700032, India}
\emailAdd{ parijat.dey@physics.uu.se, psnk2348@iacs.res.in}
\preprint{UUITP-27/21}

\abstract{In the bulk reconstruction program, one constructs boundary representations of bulk fields.  We investigate the relation between the global/Poincare and AdS-Rindler representations for AdS$_2$. We obtain the AdS-Rindler smearing function for massive and massless fields and show that the global and AdS-Rindler boundary representations are related by conformal transformations. We also use the boundary representations of creation and annihilation operators to compute the Bogoliubov transformation relating global modes to AdS-Rindler modes for both massive and massless particles.}

\begin{document}
\maketitle
\section{Introduction}

The AdS/CFT correspondence\cite{Maldacena:1997re,Gubser:1998bc,Witten:1998qj} is the conjectured equivalence between a gravitational theory in a $d+1$-dimensional asymptotically anti-de Sitter spacetime to a conformal field theory living on a $d$-dimensional flat spacetime. This means that one should in principle be able to describe bulk physics completely using the boundary theory. To be able to do this, one would need to express all the observables of the bulk theory in terms of the boundary theory. 

The AdS/CFT dictionary in its orignal form does not directly tell us how to translate all bulk observables to objects in the boundary theory. It gives us a relation between the boundary limit of correlation functions of a bulk field and the correlators of its dual operator in the boundary theory\cite{Banks:1998dd}: 
\begin{align}
&\label{dictionary}\lim_{Z \to 0} Z ^{-n\Delta} \langle \phi(Z_1,T_1,\X_1)\phi(Z_2,T_2,\X_2)\cdots\phi(Z_n,T_n,\X_n)\rangle\nn
&=  \langle 0|\mathcal{O}(T_1,\X_1)\mathcal{O}(T_2,\X_2)\cdots\mathcal{O}(T_n,\X_n)|0\rangle\,.
\end{align}
This is the extrapolate dictionary for a scalar field. Using \eqref{dictionary}, it is indeed possible to construct an operator $\phi_{cft}(Z,T,\X) $ in the boundary theory such that: 
\begin{align}
\label{brep} \langle 0|\phi_{cft} (Z_1,T_1,\X_1).. \phi_{cft}(Z_n,T_n,\X_n)|0 \rangle =   \langle \phi (Z_1,T_1,\X_1) ..\phi(Z_n,T_n,\X_n). \rangle
\end{align}
This relation says that the correlation functions of $\phi_{cft}$ computed from the boundary theory will match exactly with the bulk correlation functions of the field $\phi$, calculated using field theory in an Anti de-Sitter spacetime. So if one can construct the operator $\phi_{cft}$, one would be able to carry out all calculations of the bulk theory entirely from the boundary theory. Operators like $\phi_{cft}$ are known as boundary representations of bulk fields and their construction goes by the name of `bulk reconstruction'. Boundary representations have been constructed for different fields in different asymptotically AdS backgrounds\cite{Dobrev:1998md, Bena:1999jv,Hamilton:2005ju,Hamilton:2006az,Hamilton:2006fh,Heemskerk:2012np,Papadodimas:2012aq,Kabat:2011rz,Heemskerk:2012mn, Kabat:2012hp,Leichenauer:2013kaa,Sarkar:2014dma,Sarkar:2014jia,Guica:2014dfa,Roy:2015pga,Kabat:2016rsx,Kabat:2017mun,Kabat:2018pbj,Foit:2019nsr,Kajuri:2020bvi}.  We refer to \cite{Kajuri:2020vxf} for a recent review.

To construct a boundary representation one has to work in a given coordinate system.The boundary representation turns out to be a nonlocal operator in the boundary. The boundary representation $\phi_{cft}$ of a bulk scalar $\phi$ would be given by:
\beq 
\phi_{cft}(Z,T,\X) =  \int dT' \, d^{d-1}\X' K(Z,T,\X; T',\X') \O(T',\X')\,,
\eeq
 Here $(Z,T,\X) $ are the bulk coordinates and $(T',\X') $ are boundary coordinates. $\O(T',\X')$ is the primary operator dual to the bulk field $\phi$. $K(Z,T,\X; T',\X') $ is known as the smearing function and is given by:
 \beq \label{pme}
 K(Z,T,\X; T',\X') =  \int \, d\o \, d\k \,  A_{\o,\k} f_{\o,\k}(Z,T,\X)e^{i \o T' -i \k\cdot \X'}\,,
 \eeq 
 where $f_{\o,\k}(Z,T,\X)$ are the mode solutions to the Klein Gordon equation in the $(Z,T,\X)$ coordinate system and $ A_{\o,\k}$ is a constant. One can also obtain boundary representations of the creation and annhihilation operators :
\begin{align} \label{equiv}
a_{\o,\k}& \propto \int dT \, d^{d-1}\X \,e^{i \o T -i\k \cdot \X}\O(T,\X) \, ,\\
a^\dagger_{\o,\k} & \propto \int dT \, d^{d-1}\X \, e^{-i \o T+i\k \cdot \X}\O(T,\X)\,.
\end{align}

One can obtain boundary representations by working in different charts. For global and Poincare charts, one can obtain analytical expressions for the smearing functions and show that the global and Poincare boundary representations of fields are related via a conformal map\footnote{This equivalence is up to allowed redefinitions of the smearing function. See \cite{Hamilton:2005ju,Hamilton:2006az}}. However for AdS-Rindler coordinates in dimensions greater than two, boundary representation one obtains in this coordinate chart is not conformally equivalent to the global/Poincare representations. The integral in \eqref{pme} is known to diverge in three or higher dimensions when AdS-Rindler modes are substituted for $ f_{\o,\k}$. This means that the smearing function for the AdS-Rindler boundary representation does not exist as a function, although it can be understood either in a distributional sense\cite{Papadodimas:2012aq}. Alternately, the boundary representations can be understood in momentum space\cite{Morrison:2014jha}. Either way, the divergence of the AdS-Rindler smearing function shows that $\phi_{cft}$ obtained using the global or Poincare chart and $\phi^{Rind}_{cft}$ are inequivalent.\footnote{There is one way of obtaining boundary representations where this problem is resolved. This involves complexifying the boundary coordinates. In this approach, it has been shown that 2+1 dimension AdS-Rindler chart, one does obtain an analytic expression for a smearing function that is manifestly covariant\cite{Hamilton:2006az,Hamilton:2006fh}. However, this involves a large analytic continuation, and it is unclear if these smearing functions are well-defined in Lorentzian signature\cite{Morrison:2014jha}. In this paper, we work with a boundary field theory purely in Lorentzian signature.}

In this paper, we are interested in comparing and relating the different boundary representations in two dimensions. A major advantage for $AdS_2$ is that the smearing function is not expected to be divergent. This is because there are no evanescent modes (modes for which $\o^2-\k^2 <0$) amongst the mode solutions for a scalar field in AdS-Rindler coordinates. Evanescent modes are closely related to the divergence of the AdS/Rindler smearing function\cite{Papadodimas:2012aq,Rey:2014dpa}. Therefore divergences should not occur and the boundary representations corresponding to different charts should be equivalent. For the case of a massless scalar, these expectations are borne out by the AdS/Rindler smearing function computed in \cite{Lowe:2008ra}. In this paper, we compute the smearing functions corresponding to different charts for the case of a massive scalar and show that they're equivalent. This is our first result.

We also relate the different boundary representations of creation/annihilation operators. In effect, this constitutes a calculation of bulk Bogoliubov coefficients between global and AdS/Rindler modes purely from the boundary theory. We compute the Bogoliubov coefficients for both massive and massless fields. For massless fields, we could derive an exact expression while for massive fields we were able to express the Bogoliubov coefficients as a summation. The computation of Bogoliubov coefficients is our second result. We note that a complete calculation of the Bogoliubov coefficients between global and AdS-Rindler modes has not been performed before\footnote{In \cite{Belin:2018juv} Bogoliubov coefficients for the global zero mode expanded in an AdS/Rindler basis for AdS$_3$  had been computed. Their strategy is similar to ours, although not involving boundary representations and suitable only for the zero mode.}.

The paper is organised as follows. In the section \ref{eqivrep} we construct smearing functions for a massless scalar field in AdS$_2$ in global, Poincare and AdS-Rindler coordinates and demonstrate that \eqref{master} holds. These smearing functions have appeared before in the literature, this section has been included for completeness and to review the steps involved in bulk reconstruction in detail. In section \ref{equivmassive} the smearing function for massive fields in AdS-Rindler coordinates is derived the conformal equivalence of AdS-Rindler boundary representation with global/Poincare representations is established.  The Bogoliubov coefficients between global and AdS-Rindler modes for both massless and massive particles are calculated in section \ref{bcoeff}. We conclude in  section \ref{conc} with some future directions. The appendix \ref{normalisation} gives some of the calculational details.

%%%%%%%%%%%%%%%%%%%%%%%%%%%%%%%%%%%%%%%%%%%%%%%%%%%%%%%%%%%%%%%%%%
 
\section{Equivalence of boundary representations  in AdS$_2$ for massless fields}\label{eqivrep}

In this section, we will show that the boundary representations for a massless field constructed in global, Poincare and AdS-Rindler coordinates are all conformally equivalent.  First, we establish the relation between smearing functions obtained in two different coordinate systems when the corresponding boundary operators are related by a conformal map. Let us consider two 2d bulk coordinate systems, say AdS-Rindler $(z,\e)$ and Poincare $(Z,T)$. 
The corresponding boundary representations will be given by: 
\begin{align}
\label{rme2} \phi_{cft} (z,\e)&= \int \, d\e' \,   K(z,\e; \e') \O(t') \,,\\
\label{pme2} \phi_{cft} (Z,T) &= \int \,  dT' \,  K(Z,T; T') \O(T')\,.
\end{align}
Under a boundary conformal transformation $\e' \to T'$, the AdS-Rindler boundary representationb $\phi_{cft}(z,\e)$ transforms as:
\begin{align}
\int \, d\e' \,   K(z,\e; \e') \O(t') \to \int \, \left|\frac{d \eta' }{dT'} \right| \, dT' \,  K(z,\e; \e') \left|\frac{d \eta' }{dT'} \right| ^{-\Delta} \O(T')\,,
\end{align}
 It follows that if \eqref{rme2} and \eqref{pme2} are related via a conformal transformation, then the AdS-Rindler smearing function $K(z,\e; \e')$ and the Poincare smearing function $K(Z,T; T')$ will be related by:
\beq \label{master}
\left|\frac{d \eta' }{dT'} \right| ^{1-\Delta}K(z,\e; \e') = K(Z,T; T')\,.
\eeq
This is the relation that we will prove.

For massless fields in AdS$_2$, $\D =1$, so this requirement simply translates to showing that all the smearing functions are equal. We can get one from the other simply by a change of coordinates.

The results presented here are not novel -- although the question of equivalence between AdS-Rindler and global/Poincare boundary representations has not been considered before, the corresponding smearing functions for a massless scalar have been computed in the past. For global or Poincare coordinates, representations for the massive field were computed in \cite{Hamilton:2005ju} from which the smearing function for the massless field can be obtained by taking the massless limit. For AdS-Rindler a green function derivation of the massless case was presented in \cite{Lowe:2008ra}. The main purpose of this section is to review the steps involved in bulk reconstruction method in the simple case of massless fields. We compute smearing functions using the method of mode expansion in each case. A useful reference for field theory in AdS$_2$ is \cite{Spradlin:1999bn}.

Let us first consider the global smearing function. The AdS$_2$ metric in global coordinates is given by:
\beq \label{rindmetr}
ds^2 = \frac{1}{\cos^2 \rho}(-d \tau^2 + d \rho^2)\,,
\eeq
 where $-\p < \rho < \p $ and $- \infty < \tau < \infty$.
The free massless field satisfies the Klein Gordon equation:
\beq \label{box}
\Box \f =0\,,
\eeq
where $\Box \f=\frac{1}{\sqrt{-g}}\partial_{\mu}\left(\sqrt{-g} \,g^{\mu \nu} \partial_{\nu} \f\right)$. 
Because of $\tau$-translation symmetry, the mode solutions will be of the form $\f(\rho, \tau)=f_\o (\rho) e^{i \o \tau}$. Substituting in \eqref{box} we get the simple equation
\beq 
\frac{d^2 f_\o}{d \rho^2} +\o^2 \rho =0\,.
\eeq
The solutions are given by
\beq 
f_\o(\rho) = A_\o \cos \o \rho + B_\o \sin \o \rho\,,
\eeq
where $A_\o$ and $B_{\o}$ are $\o$-dependent constants. 
The extrapolate dictionary holds for the normalizable mode i.e the one that vanishes at the boundary. The boundary conditions are therefore that $f_\o(\rho=\pm \p)=0$. %at $\rho =-\p$ and $\rho =\frac{\pi}{2}$.
Applying the boundary conditions we get the quantized modes
\beq 
f_n(\rho) \sim \sin 2 n \rho \,,
\eeq
where $n$ is an integer. 
We will later need their boundary limit: 
\beq \label{glim}
\lim_{\rho \to \p} \frac{1}{\cos \rho} \,f_n(\rho) = - 2n \cos n\pi\,.
\eeq
So the field can be written in the mode expansion: 
\beq \label{gmode}
\f(\rho,\tau) = \sum_{n>0}\ka_n \sin 2 n \rho \,\left( e^{-2 i n \tau}a_n + e^{2 i n \tau}{a^\dagger}_n \right)\,,
\eeq
where $a_n /{a^\dagger}_n $ are the annihilation/ creation operators and $\ka_n$ is the normalization constant % The normalization constant is defined by the demand that the solutions have unit norm:
%\beq 
%\langle S_n, S_{n'}\rangle=-i  \int^{\p}_{-\p} d\rho \, S_n(\rho,\tau) \overleftrightarrow{\partial_\tau}S^*_{n'}(\rho,\tau) \bigg \rvert_{\tau=\text{constant}}  = 2 \pi \d_{n,n'}\,,
%\eeq 
%where $S_n(\rho,\tau) =N_n f_n (\rho) e^{2 i n \tau}$. Hence we get the following normalisation constant
\beq \label{gnorm}
\ka_n = \sqrt{\frac{2}{n}}.
\eeq
We will not need $\ka_n$ to compute the smearing function, but it will be used in section \ref{bcoeff} to calculate the Bogoliubov coefficients. 

Let us proceed to construct the boundary representation of the bulk fields. In global coordinates, the extrapolate dictionary reads:
\beq \label{gdictionary}
\lim_{\rho \to \p} \frac{1}{\cos \rho} \, \f (\rho,\tau) = \O(\tau)
\eeq
%In our case $\D=1$.
Substituting \eqref{gmode} in \eqref{gdictionary} and using \eqref{glim} we then get:
\beq \label{modes}
-\sum_{n>0} 2 \ka_n n \cos n\pi \,\left( e^{-2 i n \tau}a_n + e^{2 i n \tau}{a^\dagger}_n \right) = \O(\tau)
\eeq 
This gives us:
\begin{align} \label{geq}
 a_n = -\frac{1}{2 \ka_n n \cos n\pi} \O_n \,, \\ \label{geq2}
{a^\dagger}_n = -\frac{1}{2 \ka_n n \cos n \pi}\O_{-n}\,,
\end{align} 
where we have defined:
\begin{align}\label{gcrea}
\O_n =\int^\pi_{-\pi} d\tau \, \O(\tau) e^{2 i n \tau}\,, \\ \label{gann} \O_{-n} =\int^\pi_{-\pi} d\tau \, \O(\tau) e^{-2 i n \tau}\,.
\end{align}
We could have chosen any interval of $2\pi$ as the range of the above integrals. Keeping the range to be $-\pi$ to $\pi$  is convenient for comparing with other smearing functions in other coordinates.
Substituting \eqref{geq} and \eqref{geq2} in \eqref{gmode} we obtain the boundary representation of the bulk field:
\beq \label{grep}
\f(\rho,\tau) = -\sum_{n>0} \frac{1}{2 n \cos n\pi} \sin 2 n \rho \,\left( e^{-2 i n \tau}\O_n + e^{2 i n \tau}\O_{-n} \right)\,.
\eeq 
Here and afterwards we have dropped the suffix 'cft' and refer to boundary representations as $\phi$. 
Note that the normalization constant has canceled out of the expression but the boundary limit of the mode solution \eqref{glim} enters in the denominator. Using \eqref{gcrea} and \eqref{gann}, \eqref{grep} can be written in the form:
\beq 
\f(\rho,\tau) = \int_{-\infty}^{\infty} d\tau' K(\rho,\tau;\tau') \O(\tau')\,,
\eeq
 where the  smearing function $K(\rho,\tau;\tau')$ is given by: 
 \beq 
 K(\rho,\tau;\tau') =- \sum_{n>0} \frac{1}{2  n \cos n\pi} \sin 2 n \rho \, \cos 2n(\tau-\tau')\,.
 \eeq
Let us evaluate this for the case where $\tau =0$. Then we have that:
\beq 
- \sum_{n>0} \frac{1}{2  n \cos n\pi} \sin 2 n \rho \, \cos 2n\tau'= \frac{i}{8} \log \left( \frac{ 1- i e^{2i(\tau'-\rho)}}{ 1- i e^{-2i(\tau'-\rho)}} \frac{ 1- i e^{-2i(\tau'+\rho)}}{ 1- i e^{2i(\tau+\rho)}} \right)\,.
\eeq 
This is the same factor that appeared in \cite{Hamilton:2005ju} for the massive case. As had been noted in that paper, the  expression simplifies since $f(x) = -i \log \frac{1+e^{ix}}{1+ e^{-ix}}$ is a sawtooth function : $f(x)=x$ for $-\pi<x <\pi$ and $f(x+2\pi)=f(x)$. Therefore we finally have:
\beq 
K(\rho,0;\tau') =\frac{\pi}{4} \theta \left( \cos(\tau') - \cos(\p - \rho) \right)\,.
\eeq
In the general case, the expression is:
\beq \label{gsmear}
K(\rho,\tau;\tau') =\frac{\pi}{4} \theta \left( \cos(\tau -\tau') - \cos(\p - \rho) \right)\,.
\eeq
This is the smearing function in the global coordinates. 

Next, we follow the same steps to compute the smearing function in Poincare coordinates. 
Poincare coordinates are related to the global coordinates by:
\beq \label{pcord}
Z= \frac{\cos \rho}{\cos \tau + \sin \rho} \, ,\qquad T= \frac{\sin \tau}{\cos \tau + \sin \rho}\,,
\eeq
where $0<Z< \infty$ and $-\infty < T < \infty$.
The $AdS_2$ metric in Poincare coordinates is given by
\begin{align}
ds^2 &=\frac{R^2}{Z^2}\bigg(-dT^2+dZ^2\bigg) \,.
%{\rm where}  \quad 0< Z< \infty\,, \quad -\infty< T< \infty\,.
\end{align}
These coordinates cover an interval of $-\pi < \tau < \pi$ of the global chart. In these coordinates, the Klein-Gordon equation for a massless scalar will admit a solution of the form $h_\o(Z,T)=R_\o(Z)e^{i \o T}$
 where 
\begin{align}\label{pkg}
 R''_{\o}(Z)+w^2 R_\o(Z)=0\,.
\end{align}
There is a single boundary condition: the field should vanish as $Z\to 0$. The solution to \eqref{pkg} compatible with this boundary condition is $R_\o(Z)= \sin \o Z$.
This mode has the boundary limit:
\beq \label{plim}
\lim_{Z \to 0} \frac{\sin \o Z}{Z} = \o\,.
\eeq
The normalizable mode solutions are then: 
\begin{align}
h_{\o}(Z, T)=  \sin \o Z\, e^{i \o T}\,.
\end{align}
The smearing function is given by
\beq\label{poincaresm}
K(Z, 0;T') = \int_0^{\infty } d\o \,\frac{1}{\o}\sin \o Z\, \cos{ \o T'}\nn
\eeq
where the $1/\o$ factor comes from \eqref{plim}.
We then get:
\begin{align}
K(Z, 0;T')&=\frac{\pi}{4} \bigg(\text{sgn}(T'+Z)-\text{sgn}(T'-Z)\bigg)=\frac{\pi}{4} \theta \left(Z-T'\right)\,.
\end{align}
So the boundary representation of a massless bulk scalar in Poincare coordinates is given by:
\beq \label{phkll}
\phi(Z,T) = \int dT' K(Z, T;T') \O(T')\,,
\eeq
where
\begin{align}
K(Z, T;T')=\frac{\pi}{4} \theta \left(Z-\left|T-T'\right|\right)\,.
\end{align}
Expressed in terms of global coordinates using \eqref{pcord}, this is the same expression as \eqref{gsmear}.

Now we will obtain the boundary representation in AdS-Rindler coordinates and show it to be equivalent to the global boundary representation. The AdS-Rindler coordinate system is related to the global coordinates by:
\beq \label{rcord}
{z}= \frac{\cos \rho}{\cos \tau}\qquad ; \qquad \tanh \eta = \frac{\sin \tau}{\sin \rho}\,,
\eeq
where  $0< z< 1$ and $ -\infty< \eta< \infty$.
%Here $0<z<L, -\infty < \eta < \infty$. 
The $AdS_2$ metric in Rindler coordinates is given by
\begin{align}\label{adsrindler2}
ds^2 &=\frac{L^2}{z^2}\bigg(-f(z) d{\eta}^2+\frac{1}{f(z)}dz^2\bigg)\,, \qquad f(z)=1-{z^2}\,,
%{\rm where}& \quad 0< z< L\,, \quad -\infty< \eta< \infty\,.
\end{align}
This covers an interval of $-\p <\tau <\p$ on the right boundary. We will compute the smearing function for a field at a point in this right Rindler patch. The computation is along the same lines as the global and Poincare cases. 
Solving the Klein Gordon equation and imposing the boundary condition $\f(z=0, \eta)=0$ we have the normalizable modes:
\begin{align}
g_\o(z, \eta)= \int_{0}^{\infty} d\o\, \sin\left( \o \tanh^{-1}{z}\right) e^{i \o \e}\,.
\end{align}
The normalization constant for these modes turns out to be:
\beq 
\mathcal{N}_\o =\frac{1}{2\sqrt{ \o}}
\eeq
Then the bulk field has the mode expansion:
\beq 
\f(z,\eta)= \int_{0}^{\infty} \, d\o \, \mathcal{N}_\o \, g_\o(z, \eta)\, b_\o + c.c.
\eeq
We will need the boundary limit of the mode functions:
\begin{align}\label{rlim}
\lim_{z\to 0} \frac{1}{z}\sin\left( \o \tanh^{-1}{z}\right)=\o
\end{align}
Following the same steps as before, we find that the bulk creation/annihilation operators can be represented by boundary operators:
\begin{align}\label{req}
 b_\o= \frac{1}{ \mathcal{N}_\o \o} \O^{Rind}_\o \,,\\ \label{req2}
{b^\dagger}_\o =\frac{1}{\mathcal{N}_\o \o}\O^{Rind}_{-\o}
\end{align} 
where we have defined:
\begin{align} \label{rindmode}
\O^{Rind}_\o  =\int^\infty_{-\infty} d\eta\,  \O(\eta) e^{i \o \eta} \,,\\ \label{rindmode2}  \O^{Rind}_\o  =\int^\infty_{-\infty} d\eta \, \O(\eta) e^{-i \o \eta}\,.
\end{align}
The smearing function at $\eta=0$ is given by
\begin{align}\label{rindsm}
K(z, 0;\eta')&=\int_{0}^{\infty} {d\o} \frac{1}{\o}\sin\left(L \o \tanh^{-1}\frac{z}{L}\right) \cos(\o \e')\nn
%&= \frac{\pi}{4}\bigg(\text{sgn}\left(\e+L \tanh^{-1}\frac{z}{L}\right)-\text{sgn}\left(\e'-L \tanh^{-1}\frac{z}{L}\right)\bigg)\nn
&=\frac{\pi}{4} \theta \left( \tanh^{-1}{z} -\eta' \right)\nn
&=\frac{\pi}{4} \theta \left({z} -\tanh{ \eta'} \right)
\end{align}
where in the last step we have used the fact that in the range $-1<x<1$, $\tanh^{-1}x$ is single-valued (and real), so $\tanh^{-1}x=\tanh^{-1}y$ iff $x=y$.
Transforming the smearing function \eqref{rindsm} to global coordinates using \eqref{rcord} we see that it is indeed the same as the global smearing function \eqref{gsmear}. Thus we find that the AdS-Rindler boundary representation of the massless scalar field in AdS$_2$ is related to the boundary representations in global and Poincare coordinates via conformal transformations.

%%%%%%%%%%%%%%%%%%%%%%%%%%%%%%%%%%%%%%%%%%%%%%%%%%%%%%%%%%%%%%%%%%%%%%
\section{Equivalence of smearing functions in AdS$_2$ for massive fields}\label{equivmassive}

Now we turn to massive fields. We will derive the smearing function for massive fields in AdS-Rindler coordinates and prove that it is related to the global smearing function for massive fields via \eqref{master} \footnote{In \cite{Hamilton:2005ju} an AdS-Rindler representation was constructed in AdS$_2$ by simply re-writing the global smearing function in AdS/Rindler coordinates. This would be incorrect in higher dimnesions, but our result shows that this was correct in two dimensions}. The EOM for massive fields in the background \eqref{adsrindler2} reads:
\begin{align}\label{masseom}
\Box \f=M^2 \f\,.
\end{align}
As before, we assume solutions of the form:
\begin{align}\label{solmass}
 \f(z, \eta)= F_{\o}(z,\eta)= \psi_{\o}(z)e^{-i \o \eta}\,.
\end{align}
Substituting \eqref{solmass} in \eqref{masseom} gives
\begin{align}\label{eom2m}
\frac{\o^2}{1-{z^2}} \psi_{\o}(z)+\partial_z\bigg( \left(1-{z^2}\right) \,\partial_z \psi_{\o}(z)\bigg)=\frac{\D^2-\D}{z^2}  \psi_{\o}\,,
\end{align}
where we have used $M^2=\D^2-\D $. %For simplicity we will assume $L=1$. 
This equation admits two independent solutions:
\begin{align}\label{massol2}
\psi_{\o}(z)&=\mathcal{C}_{\o}\, z^{\D}(-1)^{\frac{1}{2}-\frac{i \o}{2}} (1-z^2)^{-\frac{i \o}{2}}\, _2F_1\left(\frac{1}{2} (\Delta -i \o),\frac{1}{2} (\Delta -i \o+1);\Delta +\frac{1}{2};z^2\right)\nn
&+\mathcal{D}_{\o} \,i\,(-1)^{\frac{1}{2}-\frac{i \o}{2}-\D} z^{1-\D}\,(1-z^2)^{-\frac{i \o}{2}}\, _2F_1\left(\frac{1}{2} (-\Delta -i \o+1),\frac{1}{2} (-\Delta -i \o+2);\frac{3}{2}-\Delta ;z^2\right)\,.
\end{align}
The normalizable modes are those with $z^{\D} $ fall-off near $z \sim 0$. These are given by:
\begin{align} \label{rmmode}
F_{\o}(z,\eta)= \mathcal{C}_{\o}\, z^{\D}(-1)^{\frac{1}{2}-\frac{i \o}{2}} (1-z^2)^{-\frac{i \o}{2}}\, _2F_1\left(\frac{1}{2} (\Delta -i \o),\frac{1}{2} (\Delta -i \o+1);\Delta +\frac{1}{2};z^2\right)e^{-i \o \eta}\,,
\end{align}
where $\mathcal{C}_{\o}$ is the normalisation constant given in appendix \ref{normalisation}.
The boundary limit of \ref{rmmode}  is:
\begin{align}\label{rmlim}
 \lim_{z\to 0}  z^{-\D}F_{\o}(z,\eta)= (-1)^{\frac{1}{2}-\frac{i \o}{2}} \mathcal{C}_{\o}e^{-i \o \eta}\,.
\end{align}
We now consider the smearing function for a bulk point $(z, \eta=0)$.
Using \eqref{rmmode} and \eqref{rmlim}, we arrive at the expression for the smearing function
\begin{align}
K(z,0; \eta')=\int_{-\infty}^{\infty}d\o\, z^{\D} (1-z^2)^{-\frac{i \o}{2}}\, _2F_1\left(\frac{1}{2} (\Delta -i \o),\frac{1}{2} (\Delta -i \o+1);\Delta +\frac{1}{2};z^2\right)e^{-i \o \eta'}\,.
\end{align}
At this stage, we use the following identity for the hypergeometric function
\begin{align}
\, _2F_1\left(a,a+\frac{1}{2};c;x\right)={\left(\sqrt{x}+1\right)^{-2 a}}{\, _2F_1\left(2 a,c-\frac{1}{2};2 c-1;\frac{2 \sqrt{x}}{\sqrt{x}+1}\right)}\,,
\end{align}
and use the integral representation of the same
\begin{align}
\, _2F_1\left(a,b;c;x\right)=\frac{\G(c)}{\G(b)\,\G(c-b)}\int_0^{1}dt \,t^{b-1}\,(1-t)^{c-b-1}\,(1-x\,t)^{-a}\,.
\end{align}
This results in
\begin{align}\label{smass}
&K(z,0;\eta')\nn
&=\frac{\G(2\D)}{\G^2(\D)}\left({\frac{z}{1+z}}\right)^{\D}\int_{-\infty}^{\infty}d\o \int_0^{1}dt \,e^{-i \o \eta'}  (1-z^2)^{-\frac{i \o}{2}}(1+z)^{i\o}\,(t(1-t))^{\D-1}\left(1-\frac{2t z}{1+z}\right)^{i\o-\D}\nn
&=\frac{\G(2\D)}{\G^2(\D)}\left({\frac{z}{1+z}}\right)^{\D} \int_0^{1}dt \int_{-\infty}^{\infty}d\o\,e^{i \o \left(-\eta'+\log \frac{1+z-2t z}{(1-z^2)^{\frac{1}{2}}}\right)}  \,(t(1-t))^{\D-1}\left(1-\frac{2t z}{1+z}\right)^{-\D}\nn
&=\frac{\G(2\D)}{2 \pi \G^2(\D)}\left({\frac{z}{1+z}}\right)^{\D} \int_0^{1}dt \,\delta\left(-\eta'+\log \frac{1+z-2t z}{\sqrt{1-z^2}}\right)  \,(t(1-t))^{\D-1}\left(1-\frac{2t z}{1+z}\right)^{-\D}\,,
\end{align}
where we have changed the order of the integral by first doing the $\o$ integral and then the $t$ integral. 
Now we write the dirac delta function as
\begin{align}
\delta \left(-\eta'+\log \frac{1+z-2t z}{\sqrt{1-z^2}}\right) =\bigg|\frac{\sqrt{1-z^2}}{-2 z e^{-\eta'}}\bigg|\delta\bigg(t-\frac{1+z-e^{\eta'}\sqrt{1-z^2}}{2z}\bigg)\,.
\end{align}
We write the $t$ integral as
\begin{align}\label{ft}
\int_0^{1}dt \,f(t)=\frac{1}{2}\int_{-\infty}^{\infty }dt \,f(t) \bigg(\text{sgn}\left(t)\right)-\text{sgn}\left(t-1)\right)\bigg)\,.
\end{align}
Putting it all together we get the following smearing function\footnote{The $t$ integral can be done without introducing \eqref{ft} as well. The $t$ integral in the last line of \eqref{smass} survives if 
\begin{align}
0 \leq \frac{1+z-e^{\eta'}\sqrt{1-z^2}}{2z} \leq 1
\end{align}
which implies
\begin{align}
 %&\eta \geq \frac{1}{2}\log \left(\frac{1-z}{z+1}\right)\, \qquad {\rm and }\,\,  \eta\leq \frac{1}{2}\log \left(\frac{z+1}{1-z}\right) \,\quad {\rm when} \,\,0< z<1\nn
 & \eta' \geq -\tanh ^{-1}(z)\, \qquad {\rm and }\,\quad  \eta'\leq\tanh ^{-1}(z)\,\quad {\rm when} \,\quad 0< z<1\,.
\end{align}
This gives the same answer.}:
\begin{align}
K(z,0;\eta')&=\frac{2^{\D-1}\G(\D+1/2)}{\sqrt{\pi}\G(\D)}\left(\frac{ \left(1-\sqrt{1-z^2} \cosh \eta'\right)}{z}\right)^{\Delta -1}\nn
& \times
\bigg(\text{sgn}\left(z+e^{\eta'} \sqrt{1-z^2}-1\right)+\text{sgn}\left(z-e^{\eta'} \sqrt{1-z^2}+1\right)\bigg)
\end{align}
where we have used the Legendre duplication formula: $\Gamma(a) \Gamma\left(a + 1/2\right) = 2^{1-2a} \; \sqrt{\pi} \; \Gamma(2a)$. Simplifying the above, we derive the final expression for the AdS-Rindler smearing function for a bulk point at $(z,\eta=0)$:
\begin{align}\label{rmsmear}
K(z,0;\eta')&=\frac{2^{\D-1}\G(\D+1/2)}{\sqrt{\pi}\G(\D)}\left(\frac{ \left(1-\sqrt{1-z^2} \cosh \eta'\right)}{z}\right)^{\Delta -1}
\theta\left(\tanh ^{-1}z - \e'\right)
\end{align}
Now we will show that \eqref{master} is satisfied for the global and AdS-Rindler representations. The global smearing function was calculated in \cite{Hamilton:2005ju}:
\beq \label{gmsmear}
K(\rho,0;\tau') = \frac{2^{\D-1}\G(\D+1/2)}{\sqrt{\pi}\G(\D)} \left(\frac{ \cos(\tau') -\sin \rho}{\cos \rho}\right)^{\D-1} \theta\left (\p-\rho - \tau \right)\,.
\eeq
We have already shown how the arguments of the theta functions in \eqref{rmsmear} and \eqref{gmsmear} in match in the massless case. It remains to match the coordinate dependent prefactors. The boundary Jacobian is:
\beq \label{J}
J = \frac{d}{d \tau'} \left(\tanh^{-1} \sin \tau' \right)= \sec \tau'\,.
\eeq
We see, using \eqref{rcord}:
\begin{align} 
 ( \sec \tau')^{\D-1} \left(\frac{ \cos(\tau') -\sin \rho}{\cos \rho}\right)^{\D-1} &= ( \cosh \eta')^{\D-1} \left(\frac{\sech(\eta
   )-\sqrt{1-z^2}}{z}\right)^{\Delta -1}
   \\&=\left(\frac{ \left(1-\sqrt{1-z^2} \cosh \eta'\right)}{z}\right)^{\Delta -1}\,,
\end{align}
which matches exactly with the coordinate dependent prefactors in \ref{gmsmear}. Thus the smearing functions are related by \eqref{master} which shows that the global and AdS-Rindler boundary operators are related by a conformal transformation.

%%%%%%%%%%%%%%%%%%%%%%%%%%%%%%%%%%%%%%%%%%%%%%%%%%%%%%%%%%%%%%%%%%%
\section{Bogoliubov Coefficients between global and AdS-Rindler modes}\label{bcoeff}

We will now calculate the Bogoliubov coefficients expressing the global annihilation and creation operators in terms of their AdS-Rindler counterparts. We will carry out the calculation using the boundary representations defined via \eqref{gcrea},\eqref{gann} and \eqref{rindmode},\eqref{rindmode2}. This is an example of a bulk calculation that is easier to carry out using boundary representations. A similar strategy had been utilized in \cite{Belin:2018juv} where Bogoliubov coefficients between global and AdS-Rindler modes in AdS$_3$ were investigated. There the coefficients were only computed for the global zero mode. Here we have computed the coefficients for all the modes.

\begin{figure}
 \centering
  \includegraphics[width=.5\linewidth]{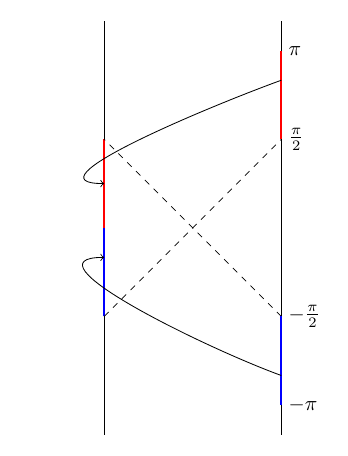}
  \caption{In this figure it is shown that how one may map the support of the global boundary representation of annihilation/creation operators to the support of the left and right AdS/Rindler representations. This is done using the antipidal map. This maps the right boundary region between $\p$ and $\pi$ to the red region in the left boundary (between $0$ and $\p$). The right boundary region shown in blue (region between $-p$ and $-\pi$)is likewise mapped to the region shown in blue in the left boundary (between $0$ and $-\p$).}
  \label{figurr}
  \end{figure}

To relate the operators $\O_n$ to the operators $\O_\o$, one seems to face an immediate obstacle. This comes from the fact that while $\O_n$ are integrated over the range $-\pi$ to $\pi$ in the global time $\tau$, $\O_\o$ are smeared over $-\pi/2$ to $\pi/2$ in $\tau$. 
To resolve this we note that $\O_n$ can be written as a sum of two operators on the two boundaries\cite{Hamilton:2005ju}. For this we need to use the antipodal mapping:
\beq 
\O^L(\tau) = (-1)^\D \O^R(\tau + \pi)\,.
\eeq
Let us tackle the massless case first. Then we have :
\beq \label{leftright}
\O_n = \O^R_n - \O^L_n\,,
\eeq
where 
\begin{align}\label{gcrea2}
\O_n =\int^\p_{-\p} d\tau \, \O(\tau) e^{i n \tau} \,,\\ \label{gann2} \O_{-n} =\int^\p_{-\p} d\tau \, \O(\tau) e^{-i n \tau}\,.
\end{align}
The mapping of $\O_n$ to $\O^R_n - \O^L_n$ is shown in figure \ref{figurr}.
The boundary support of these operators is the same as that of \eqref{rindmode} and \eqref{rindmode2}. Now the idea is to relate these two sets of operators. Then if we have that: 
\begin{align} \label{rright}
\O^R_n = A^R(n,\o) {\O_\o}^{R,Rind} +B^R(n,\o) {\O_{-\o}}^{R,Rind} \,,\\
\O^L_n = A^L(n,\o) {\O_\o}^{L,Rind} +B^L(n,\o) {\O_{-\o}}^{L,Rind}\,,
\end{align}
we can use \eqref{leftright} to write:
\beq 
\O_n =A^R(n,\o) {\O_\o}^{R,Rind} +B^R(n,\o) {\O_{-\o}}^{R,Rind}-A^L(n,\o) {\O_\o}^{L,Rind} -B^L(n,\o) {\O_{-\o}}^{L,Rind}\,.
\eeq
Further, using \eqref{geq},\eqref{geq2} we can relate $a_n$ and $a^\dagger_n$ to the operators $\O_n$ and $\O_{-n}$ respectively, and similarly we can use \eqref{req},\eqref{req2} to relate $b_\o$ and $b^\dagger_\o$ to $\O_\o$ and $\O_{-\o}$. This would give us an expression of the form:
\beq 
a_n ={\alpha^R(n,\o)}{b_\o^{R,Rind}} +{\beta^R(n,\o)} {b_\o ^{\dagger \,R,Rind}}_\o + {\alpha^L(n,\o)} {b_\o ^{L,Rind}} +{\beta^L(n,\o)} {b_\o ^{\dagger \,L,Rind}}\,,
\eeq
where $\alpha^{L(R)} (n,\o), \beta^{L(R)} (n,\o) $ are the Bogoliubov coefficients which appear in the expansion of creation/annihilation operators for the global modes in terms of the creation/annihilation operators of the AdS-Rindler modes.\footnote{Note that we differ from most literature in our convention for naming the Bogoliubov coefficients. It is more usual to refer to the coefficients appearing in the linear expansion of the modes as $\alpha, \beta$. What we call $\alpha$ and $\beta$ will be $\alpha^*$ and $-\beta^*$ in that convention.} The only unknowns are therefore $A^{L(R)} (n,\o), B^{L(R)} (n,\o) $.

We now restrict to the right AdS-Rindler wedge and  compute $A^{R}(n,\o),B^{R}(n,\o)$. 
First, we note that for $\D=1$ we have:
\beq \label{exp}
\O^R_n = \int d\tau\, \O(\tau)\,e^{i n \tau} =\int d\eta \, \O(\eta)e^{i n \tau(\e)}  =  \int d\eta \, \O(\eta) \int^\infty_{-\infty} d\o \, C(n,\o) e^{i \o \eta} \,.
\eeq 
In the first line, the factors of the Jacobian coming from the integration measure and the conformal transformation of $\O(\tau)$ have cancelled among themselves. We have taken a Fourier expansion in the last line where $C(n,\o)$ is the inverse Fourier transform:
\beq 
C(n,\o) = \int d\eta \, e^{i n \tau(\eta)} e^{-i \o \eta} \, \quad {\rm where} \qquad \tau(\eta)=\sin^{-1}\left(\tanh( \eta)\right)\,.
\eeq
Restricting $\o >0$ and defining 
\begin{align}
C(\o,n) =A(n,\o)\,, \\
C(-\o,n) =B(n,\o)\,, 
\end{align}
we obtain \eqref{rright} from \eqref{exp}. So we see that to derive the expression for $A^{R}(n,\o),B^{R}(n,\o)$ we simply need to compute the inverse Fourier transform of $e^{i \o \tau (t)}$ in terms of AdS-Rindler boundary time $\eta$.

Now we proceed to calculate $C(\o,n)$. First we recall that only even modes $n = 2m$ appeared in the mode expansion. We then get:
\begin{align}
C(2m, \o)&= \int_0^{\infty} d\e\,e^{2 i m \tau(\eta)} e^{-i w \e} \nn
&={4i \pi m }e^{-\frac{i  \pi (2m + i \o)}{2 }} \csch \left( \pi \o \right) \, _2F_1\left(1-2 m,1-i \o ;2;2\right)\,.
\end{align}
This results in:
\begin{align}
A(2m,\o) ={4i \pi m }e^{-\frac{i  \pi (2m  + i \o)}{2 }} \csch \left(  \pi \o \right) \, _2F_1\left(1-2 m,1-i \o;2;2\right)\,,\\
B(2m,\o)=-{4i \pi m }e^{-\frac{i  \pi (2m  - i \o)}{2 }} \csch \left( \pi \o \right) \, _2F_1\left(1-2 m,1+i \o ;2;2\right)\,.
\end{align}
We can now use $\eqref{geq}$ and $\eqref{req}$ to compute the Bogoliubov coefficients. The final result is:
\begin{align}
\alpha^{R} (2m,\o)=4i \pi  m \sqrt{\frac{2m}{\o}} e^{-\frac{i  \pi (2m  + i \o)}{2 }} \csch \left( \pi \o \right) \, _2F_1\left(1-2 m,1-i \o ;2;2\right)\,,\\
\beta^{R} (2m,\o)=-4i \pi  m \sqrt{\frac{2m}{\o}}e^{-\frac{i  \pi (2m  - i \o)}{2 }} \csch \left( \pi \o \right) \, _2F_1\left(1-2 m,1+i \o ;2;2\right)\,.
\end{align}
These are the Bogoliubov coefficients for expanding massless particle states in the global Fock basis in terms of the basis states of the AdS-Rindler Fock space.

Now we turn to the massive case. %We consider the case where $\D$ is an integer.  
The steps of the calculation are the same as before. 
We now have:
\beq \label{mexp}
\O^R_n = \int d\tau\ \O(\tau)\,e^{i n \tau} =\int d\eta \,J^{\D-1} \O(\eta)\,e^{i n \tau(\e)}  =  \int d\eta\, \O(\eta) \int^\infty_{-\infty} d\o\, D(n,\o) e^{i \o \eta} \,,
\eeq 
where $J$ is the Jacobian. Here
\beq 
D(n,\o) = \int_0^\infty \,d\eta\, \cosh ^{\D-1}\eta\,  e^{i n \sin^{-1}\left(\tanh( \eta)\right)} e^{-i \o \eta}\,.
\eeq
We have been able to express $D(n,\o)$ as a summation. 
\begin{align}\label{bog}
D(n, \o)=-\sum_{q=0}^{\infty}  2^{3-\Delta } \pi (-i)^{2 n} n\,  \binom{\Delta -1}{q} & e^{-\frac{i \pi  (-\Delta +2 q+i\o+1)}{2}} \csc \left(\pi   (-\Delta +2 q+i\o+1)\right)\nn & \times \, _2F_1\left(1-2 n,\left(\Delta -2 q-i\o-1\right)+1;2;2\right)
\end{align}
The normalizations for massive global and AdS-Rindler modes have been worked out in the appendix \ref{normalisation}. Here we state the results from \ref{globnorm} and  \ref{rindnorm1}:
\beq
N_n = \sqrt{\frac{4^{\Delta -1}n! \Gamma (\Delta )^2 (\Delta +n)
   }{\pi  {\o_n} \Gamma (n+2 \Delta )}}\,,
\eeq
%Normalization constant for massive AdS-Rindler modes \ref{rindnorm1}:
\beq
\mathcal{C}_{\o} = \sqrt{\frac{\pi ^2 \o \Gamma
   (\Delta -i \o) \Gamma (i \o+\Delta )}{2^{2 \Delta -5} \Gamma \left(\Delta +\frac{1}{2}\right)^2
   (\coth (\pi  \o)+1) (\pi  \o \coth (\pi  \o)+1)}}\,.
\eeq 
The boundary limit for massive AdS-Rindler modes is given by \eqref{rmlim}. The boundary limit for the massive global mode is worked out in the appendix to be: 
 \beq 
b_n= \frac{\sqrt{\pi } \Gamma \left(\Delta +\frac{1}{2}\right)}{\Gamma
   \left(\frac{1}{2}-\frac{n}{2}\right) \Gamma
   \left(\frac{n}{2}+\Delta +\frac{1}{2}\right)}\,.
  \eeq
  The Bogoliubov coefficients for the expansion of massive global modes in terms of massive AdS-Rindler modes can then be written as:
 \begin{align}
 \alpha^R(n,\o) =\frac{(-1)^{\frac{1}{2}-\frac{i \o}{2}} \mathcal{C}_\o}{b_n N_n} D(n,\o)\,,\\
 \beta^R(n,\o) =  \frac{(-1)^{\frac{1}{2}-\frac{i \o}{2}} \mathcal{C}_\o}{b_n N_n}  D(n,-\o)\,.
 \end{align}

%%%%%%%%%%%%%%%%%%%%%%%%%%%%%%%%%%%%%%%%%%%%%%%%%%%%%%%%%%%%%%%%%%%%%%

%%%%%%%%%%%%%%%%%%%%%%%%%%%%%%%%%%%%%%%%%%%%%%%%%%%%%%%%%%%%%%%%%%%%%%
\section{Discussions}\label{conc}

We studied the relation between the global and AdS-Rindler boundary representations for fields and creation/annihilation operators in AdS$_2$.  We found that -- unlike in higher dimensions -- the AdS-Rindler smearing in AdS$_2$ function does not diverge. We derived the AdS-Rindler smearing function in AdS$_2$ and showed that the AdS-Rindler and global representations are related by a conformal transformation. 

We also related the boundary representations of the annihilation and creation operators for the global and AdS-Rindler modes. This allowed us to compute Bogoliubov coefficients between the global and AdS-Rindler modes. We were able to express all global modes in terms of AdS-Rindler modes. For massless fields, we obtained an exact analytic expression while for massive fields we were able to express the coefficients in terms of a summation. 

One interesting consequence of our work is that the representations corresponding to overlapping AdS/Rindler wedges are conformally related since each is conformally related to the global representation. A general proof of conformal equivalence of different AdS/Rindler representations was given in  \cite{Kim:2016ipt,Kajuri:2021vkg}, here we have confirmed it for AdS$_2$. It should be noted that this does not contradict the proposal of \cite{Almheiri:2014lwa} that boundary representations for overlapping wedges should only agree in the code subspace. The paradox which motivated the code subspace proposal does not arise in $AdS_2$ where the boundary cannot be divided into different subregions.\footnote{In the original draft we had mistakenly suggested a possible contradiction. We thank Jung-Wook Kim for pointing out this error.}

It will be interesting to see if the computation of Bogoliubov coefficients can be carried out in higher dimensions.

\section*{Acknowledgments}
%%%%%%%%%%%%%%%%%%%%%%%%%%
NK would like to thank Jung-Wook Kim for helpful discussions. 
The research of PD is supported by the Knut and Alice Wallenberg Foundation grant KAW 2016.0129 and the VR grant 2018-04438.

%%%%%%%%%%%%%%%%%%%%%%%%%%%%%%%%%%%%%%%%%%%%%%%%%%%%%%%%%%%%%%%%%%%
\appendix
\section{Normalisation}\label{normalisation}
The massive scalar field of mass $M$ in the global background \ref{rindmetr} can be written as 
\begin{align}
\f(\tau, \rho)&= \sum_{n>0} N_n  e^{-i \o_n \tau} \cos^{2\D} \rho\, \,C^{\D}_{n}(\sin \rho)+ {\rm c.c.}
%\f_n&=N_n  e^{-i w_n \tau} \cos^{2\D} \rho\, C^{\D}_{n}(\sin \rho)\,,\nn
\end{align}
where $w_n=\D+n=\frac{1}{2}+\sqrt{\frac{1}{4}+M^2}$ and $C^{\D}_{n}(x)$ is the Gegenbauer polynomial. $N_n$ is the normalisation constant.
The wave function is normalised when
\begin{align}
\langle \f_n | \f_{n'}\rangle= \int_{-\pi/2}^{\pi/2} d\rho \sqrt{-g}\,g^{\tau \tau} \f_n(\rho, \tau)^*\overleftrightarrow{\partial_t }\f_{n'}(\rho,\tau)\bigg|_{\tau={\rm{constant}}}=2\pi\delta_{n, n'}\,,
\end{align}
where 
\begin{align}
\f_n= N_n  \cos^{2\D} \rho\, C^{\D}_{n}(\sin \rho)\,.
\end{align}
This results  the following normalisation constant for the massive global modes
\begin{align}\label{globnorm}
N_n=\sqrt{\frac{4^{\D}n! (n+\D)\G^2(\D)}{2 \o_n \G(n+2\D)}}\,.
\end{align}

Now we compute the normalisation for the massive AdS-Rindler modes \ref{rmmode} which should satisfy
\begin{align}\label{rindnorm}
\langle F_{\o}(z,\e), F_{\o}(z,\e)\rangle=\int_{0}^{1} dz \sqrt{-g}\,g^{\e \e} F_{\o}(z,\e)^*\overleftrightarrow{\partial_{\e} }F_{\o}(z,\e)\bigg|_{\e={\rm{constant}}}=2\pi\,.
\end{align}
As a first step to compute the integral we use the following identity for  one of the hypergeometric functions in $F_{\o}(z,\e)$
\begin{align}
_2F_1\left(a,b;c;z\right)=(1-z)^{-a} \, _2F_1\left(a,c-b;c;\frac{z}{z-1}\right)\,.
\end{align}
Next we use the Mellin-Barnes representation of the hypergeometric function
\begin{align}
_2F_1\left(a,b;c;z\right)=\frac{\Gamma (c)}{2 \pi  i \Gamma (a) \Gamma (b)}\int_{-i \infty}^{i \infty} dt \frac{ (\Gamma (-t) \Gamma (a+t) \Gamma (b+t))}{ \Gamma (c+t)} (-z)^t\,,
\end{align}
such that the norm \ref{rindnorm} is given by the triple integral, schematically
\begin{align}
\langle F_{\o}(z,\e)| F_{\o}(z,\e)\rangle=\int_{0}^{1} dz \int_{-i \infty}^{i \infty} dt  \int_{-i \infty}^{i \infty} ds (\cdots)\,.
\end{align}
We can now exchange the order of the integrals and do the $z$-integral first. Then we do the $s$ and $t$-integrals  by closing the contour on the right and evaluating the residue of the poles. This results in
\begin{align}\label{rindnorm1}
\mathcal{C}_{\o}=\sqrt{\frac{\pi ^3 \o \Gamma (\Delta -i \o) \Gamma (i \o+\Delta )}
{ 2^{2 \Delta -6} \Gamma \left(\Delta +\frac{1}{2}\right)^2 (\coth (\pi  \o)+1) (\pi  \o \coth (\pi  \o)+1)}}\,.
\end{align}
%%%%%%%%%%%%%%%%%%%%%%%%%%

%%%%%%%%%%%%%%%%%%%%%%%%%%%%%%%%%%%%%%%%%%%%%%%%%%%%%%%%%%%%%%%%%%%%%%%
\bibliographystyle{JHEP}
\bibliography{adsrindlervjhep}

\end{document}